\documentclass[prb,twocolumn,superscriptaddress,showpacs]{revtex4-1}

\pdfoutput=1

\usepackage{times}
\usepackage{amsfonts}
\usepackage{graphicx}
\usepackage{graphics}
\usepackage{amssymb}
\usepackage{layout}
\usepackage{verbatim}
\usepackage{amsfonts,epsfig}
\usepackage{bm}
\usepackage{amsmath}
\usepackage{graphicx}
\usepackage{hyperref}
\usepackage{epsf}
\usepackage{bm}
\usepackage{verbatim}

\newcommand{\ket}[1]{\left|#1\right>}
\newcommand{\bra}[1]{\left< #1 \right|}
\newcommand{\up}{\uparrow}

\begin{document}
\title{Momentum relaxation in a semiconductor proximity-coupled to a disordered s-wave superconductor:  effect of scattering on topological superconductivity}
\author{Roman M.~Lutchyn}
\affiliation{Station Q, Microsoft Research, Santa Barbara, CA 93106-6105}
\author{Tudor D. Stanescu}
\affiliation{Department of Physics, West Virginia University, Morgantown, WV 26506}
\author{S. Das Sarma}
\affiliation{Condensed Matter Theory Center, Department of Physics, University of
Maryland, College Park, MD 20742}

\date{compiled \today }

\begin{abstract}

We study the superconducting proximity effect between a conventional semiconductor and a disordered s-wave superconductor. We calculate the effective momentum relaxation rate in the semiconductor due to processes involving electron tunneling into a disordered superconductor and scattering off impurities. The magnitude of the effective disorder scattering rate is important for understanding the stability of the topological (chiral p-wave) superconducting state that emerges in the semiconductor, since disorder scattering has a detrimental effect and can drive the system into a non-topological state. We find that the effective impurity scattering rate involves higher-order tunneling processes and is suppressed due to the destructive quantum interference of quasi-particle and quasi-hole trajectories. We show that, despite the fact that both the proximity-induced gap and the effective impurity scattering rate depend on interface transparency, there is a large parameter regime where the topological superconducting phase is robust against disorder in the superconductor. Thus, we establish that the static disorder in the superconductor does not suppress the proximity induced topological superconductivity in the semiconductor.
\end{abstract}

\pacs{03.67.Lx, 71.10.Pm, 74.45.+c}

\maketitle

\begin{figure}[tbp]
\begin{center}
\includegraphics[width=0.48\textwidth]{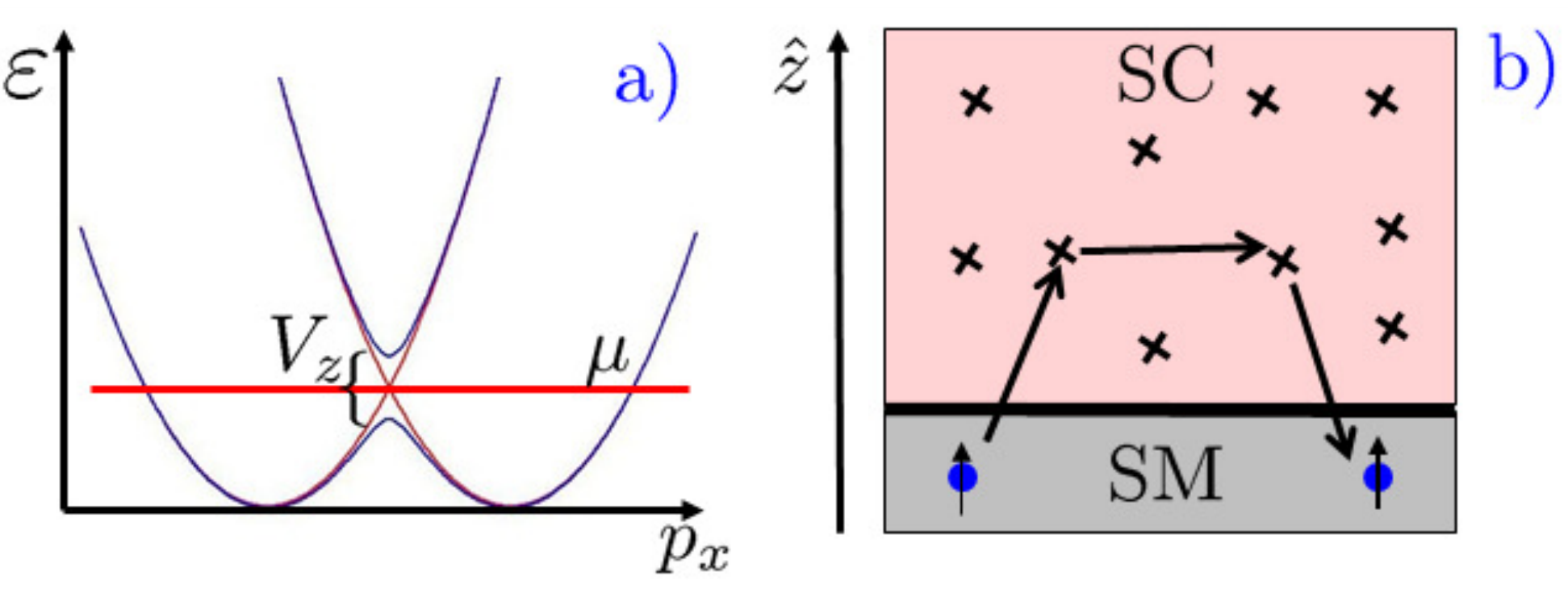}
\vspace{-7mm}
\end{center}
\caption{(Color online) a) Energy spectrum of the semiconductor with Rashba spin-orbit coupling and Zeeman splitting. The position of the chemical potential corresponding to the helical state. b) Tunneling of electrons into the superconductor leads to a momentum relaxation rate in the semiconductor $\Gamma$.} \label{fig:system}
\end{figure}

{\it Introduction.} The possibility of engineering Hamiltonians that exploit the properties of the interface between two different materials has recently attracted a lot of attention. There are many proposals that exploit magnetic, superconducting and other types of properties  for spintronics and quantum information purposes~\cite{Kane_review, Qi_review, Franz'10, levi'11}. In particular, the prospect of realizing exotic topological chiral p-wave superconducting states carrying Majorana fermions at the interface between a semiconductor with Rashba spin-orbit coupling and a conventional s-wave superconductor in sandwich structures~\cite{Sau'10, Alicea_PRB10, Lutchyn2010, Oreg2010, Potter_PRL'10, Duckheim'11, SC_Zhang'11, Zhang'11} is very intriguing. The basic concept underlying these proposals is that electrons tunneling between different materials inherit their physical properties. For example, electrons virtually propagating in the superconductor ``feel" superconducting correlations. This is the basic idea underlying the superconducting proximity effect which is used for realizing the topological $p+ip$ superconducting state at the interface. In most of the previous studies~\cite{Sau'10, Alicea_PRB10, Lutchyn2010, Oreg2010, Potter_PRL'10, Duckheim'11, SC_Zhang'11, Zhang'11}, the s-wave superconductor has been considered in the clean limit, and the effect of the superconductor disorder on the induced state was not addressed. However, it is well-known that impurity scattering in the active system (i.e. in the semiconductor) is detrimental for topological superconductivity~\cite{Motrunich'01, Gruzberg'05, Brouwer'11, Lutchyn2011, Potter'11, Stanescu'11}. While semiconductors can be grown very clean, most ordinary s-wave superconductors (e.g. Al or Nb) are disordered and have very short mean free path $l$. This motivates us to revisit the basics of the superconducting proximity effect and take into account the effects of disorder in the superconductor. As pointed out in Ref.~\onlinecite{Potter'11}, superconducting disorder might act similar to impurities in the semiconductor, see Fig.\ref{fig:system}b. Thus, the effect of superconducting disorder on the stability of the topological phase is an important open question which we investigate in this paper. The issue is of both conceptual and practical importance: On the conceptual side it may appear that disorder residing in the superconductor could be detrimental to the semiconductor superconductivity since Anderson theorem ruling out the immunity of s-wave superconductivity to non-magnetic disorder may not necessarily extend to topological superconductivity where time-reversal invariance may be explicitly broken (e.g. by an external magnetic field creating the Zeeman spin splitting), and on the practical side, disorder in the superconductor, if it turns out to be detrimental to the proximity-induced superconductivity in the semiconductor, may simply completely destroy the topological phase. We note that what is important here is the relative magnitude of the proximity-induced superconductivity in the semiconductor compared with the proximity-induced momentum relaxation rate.

The robustness of topological superconducting phases against disorder has been investigated within the simple model of one-dimensional spinless p-wave superconductors hosting Majorana zero-energy modes at the ends~\onlinecite{Motrunich'01, Brouwer'01, Gruzberg'05}. The presence or absence of these exponentially localized Majorana modes defines a topological or non-topological phase. The phase boundary between topological and non-topological phases is approximately given by $\Delta_{\rm ind} \tau \!\sim\! 1$ where $\Delta_{\rm ind}$ and $\tau$ are the proximity-induced gap in this spinless p-wave model and the impurity scattering time, respectively. In this paper we assume that the semiconductor is clean (we refer the reader to Refs.~\onlinecite{Motrunich'01, Brouwer'01, Gruzberg'05, Brouwer'11, Lutchyn2011, Potter'11} for more details on how disorder in the semiconductor affects topological superconductivity in the sandwich structures) and, thus, $\tau$ is entirely determined by the disorder scattering in the superconductor. One can ask the following question: is it possible to realize a topological phase in the sandwich structures~\cite{Sau'10, Alicea_PRB10, Lutchyn2010, Oreg2010, Brouwer'11, SC_Zhang'11, Zhang'11}, given that both the proximity-induced gap and the disorder scattering rate in the semiconductor induced by the superconductor impurities depend on the interface transparency ? This question is particularly significant in view of the topological superconducting phase in the semiconductor being equivalent to a spinless p-wave superconductor.

In this paper, we consider a simple model for the semiconductor/superconductor heterostructure and calculate the momentum relaxation rate due to electron tunneling into a disordered superconductor.
We find that the impurity scattering rate $\tau^{-1}$ in the superconductor is quite small (i.e. much smaller than proximity-induced gap $\Delta_{\rm ind}$). The reason for that is two-fold: first, the scattering rate $\tau^{-1}$ vanishes in the lowest order perturbation theory in tunneling $t$ and involves only higher-order processes; second, it is further suppressed due to the destructive quantum interference between quasi-particle and quasi-hole trajectories by a factor $1/p_F\bar\xi$, with $p_F$ and $\bar\xi$ being the Fermi momentum and coherence length in the disordered  superconductor, respectively. The importance of quantum interference effects for higher-order tunneling processes has been previously discussed in the literature in the context of two-electron tunneling~\cite{Pothier'94, Hekking'93,Hekking'94}. Therefore, we conclude that the condition for the existence of topological superconductivity can be satisfied even in the presence of substantial superconducting disorder~\cite{footnote}. In the rest of the paper, we present a detailed calculation supporting this conclusion. We emphasize that, in addition to being important for the realization of topological superconducting phases hosting Majorana fermions, our result is also very general and applies to all other superconducting heterostructures involving proximity effect.

{\it Theoretical model.} We consider a two-dimensional semiconductor in the proximity to an s-wave superconductor as shown in Fig.~\ref{fig:system}. The Hamiltonian for the semiconductor reads ($\hbar=1$)
\begin{align}\label{eq:Hamiltonian}
\!H_{\rm SM}\!=\!\int_S\!\! d^2 {\bm r}\!\Psi^\dag(\bm r)\!\!\left[\frac{\bm \hat p^2}{2m^*}\!-\!\mu\!+\!V_z \sigma_z\!+\!\alpha_R\!(\sigma_x \hat p_y \!-\! \sigma_y \hat p_x )\right]\!\!\Psi(\bm r)
\end{align}
where $S$ is 2D area of the semiconductor, $m^*$ is its effective mass, $\mu$ is the chemical potential, $\alpha_R$
is the Rashba spin-orbit coupling strength, $\Psi(\bm r)\equiv (\psi_{\up},\psi_{\downarrow})$  and $\sigma_i$ are Pauli matrices acting on the spin degree of freedom. The Zeeman splitting $V_z$ can be proximity-induced due to the presence of a ferromagnetic insulator (not shown in Fig.~\ref{fig:system}b) in 2D proposals~\cite{Sau'10} or due to an applied in-plane magnetic field $V_z=g_{\rm SM}\mu_B B$ with $g_{\rm SM}$ being the $g$-factor in the semiconductor, see Refs.~\onlinecite{Sau'10, Alicea_PRB10, Lutchyn2010, Oreg2010, Sau2010} for more details on the relevance of Eq.~\eqref{eq:Hamiltonian} for generating topological superconductivity in generic semiconductor-superconductor heterostructures.

The superconductor can be described by the BCS model with $H_{S}$ being the corresponding mean field Hamiltonian. To include disorder effects it is convenient to use exact eigenstates formalism~\cite{deGennes, Ambegaokar'92}. In the normal state, single-particle energies $\varepsilon_{n}$ and wavefunctions $\phi_n(\mathbf{x})$ in the superconductor are defined by the following one-body Schr\"{o}dinger equation:
\begin{align}\label{eq:exact}
\left[-\frac{\hbar^2}{2m}\vec{\nabla}^2+\mathcal{V}(\mathbf{x})\right]\phi_n(\mathbf{x})=\varepsilon_{n}\phi_n(\mathbf{x}),
\end{align}
where $\mathcal{V}(\mathbf{x})$ represents a particular realization of the disorder potential. We assume here that the terms breaking time reversal symmetry in the superconductor are small, {\it but not in the semiconductor,} either due to the large difference in the $g$-factors in semiconductor and superconductor~\cite{Alicea_PRB10, Lutchyn2010, Oreg2010} or because Zeeman splitting in the semiconductor is proximity-induced by ferromagnetic insulator and the amplitude for tunneling of electrons from superconductor to ferromagnetic insulator~\cite{Sau'10, Sau2010} is small and can be neglected. In this case, mean field BCS Hamiltonian can be diagonalized using the following Bogoliubov transformation:
\begin{eqnarray}\label{particleconserving}
\gamma^{\dag}_{n
\sigma}&=&\int\!{d^3\mathbf{x}}\left[U_{n}(\mathbf{x})\psi^{\dag}_{\sigma}(\mathbf{x})-\sigma
V_{n}(\mathbf{x})\psi_{-\sigma}(\mathbf{x})\right]
\end{eqnarray}
Here the transformation coefficients
$U_n(\mathbf{x})$ and $V_n(\mathbf{x})$ are given by the corresponding solution
of Bogoliubov-de Gennes equation. For spatially homogenous
superconducting gap $\Delta_0$, the functions $U_{n}(\mathbf{x})$
and $V_{n}(\mathbf{x})$ can be written as
$U_{n}(\mathbf{x})=u_n\phi_n(\mathbf{x})$ and
$V_{n}(\mathbf{x})=v_n\phi_n(\mathbf{x})$. The coherence factors
$u_n$ and $v_n$ are given by
\begin{eqnarray}
u_n^2=\frac{1}{2}\left(1\!+\!\frac{\varepsilon_{n}}{E_n}\right)
\mbox{ and }
v_n^2=\frac{1}{2}\left(1\!-\!\frac{\varepsilon_{n}}{E_n}\right).\nonumber
\end{eqnarray}
Here $E_n=\sqrt{\varepsilon_{n}^2+\Delta_0^2}$; $\varepsilon_{n}$
and $\phi_n(\mathbf{x})$ are exact eigenvalues and eigenfunctions
of the single-particle Hamiltonian~\eqref{eq:exact} which can be chosen to be real.

The tunneling Hamiltonian between semiconductor and superconductor reads
\begin{eqnarray}\label{tunneling}
H_t=\sum_{\sigma}\int{d^3\mathbf{x}d^2\mathbf{r'}}\left(T(\mathbf{x},\mathbf{r'})\psi^{\dag}_{\sigma}(\mathbf{x})\psi_{\sigma}(\mathbf{r'})\!+\!{\rm{H.c.}}\right),
\end{eqnarray}
where $\mathbf{x}$ and $\mathbf{r'}$ denote the coordinates in the
superconductor and semiconductor, respectively, and $T(\mathbf{x},\mathbf{r'})$,
in the limit of a barrier with low transparency, is defined as
\begin{align}\label{eq:tunn}
T(\mathbf{x},\mathbf{r'})\!=\!t\delta^2(\mathbf{r}\!-\!\mathbf{r'})\delta(z)\frac{\partial}{\partial z},
\end{align}
see Refs.~\onlinecite{PradaSols,
Houzet} for details.

{\it Momentum relaxation rate.} We now calculate the scattering rate of an electron in the initial state $\ket{p,\sigma}$ where $p$, $\sigma$ are electron momentum and spin, respectively, into a state $\ket{p',\sigma}$. Due to the proximity to the disordered superconductor, momentum in the semiconductor is not a good quantum number anymore and, as a result, levels $\ket{p,\sigma}$ will have some broadening $\Gamma$. Since the superconductor is a good metal, its disorder can be well-approximated by short-range impurity scattering. Without any loss of generality, we first study the case when $\alpha_R$ and $V_z$ are zero and then generalize our results to the helical regime at the end of the paper.

\begin{figure}[tbp]
\begin{center}
\includegraphics[width=0.45\textwidth]{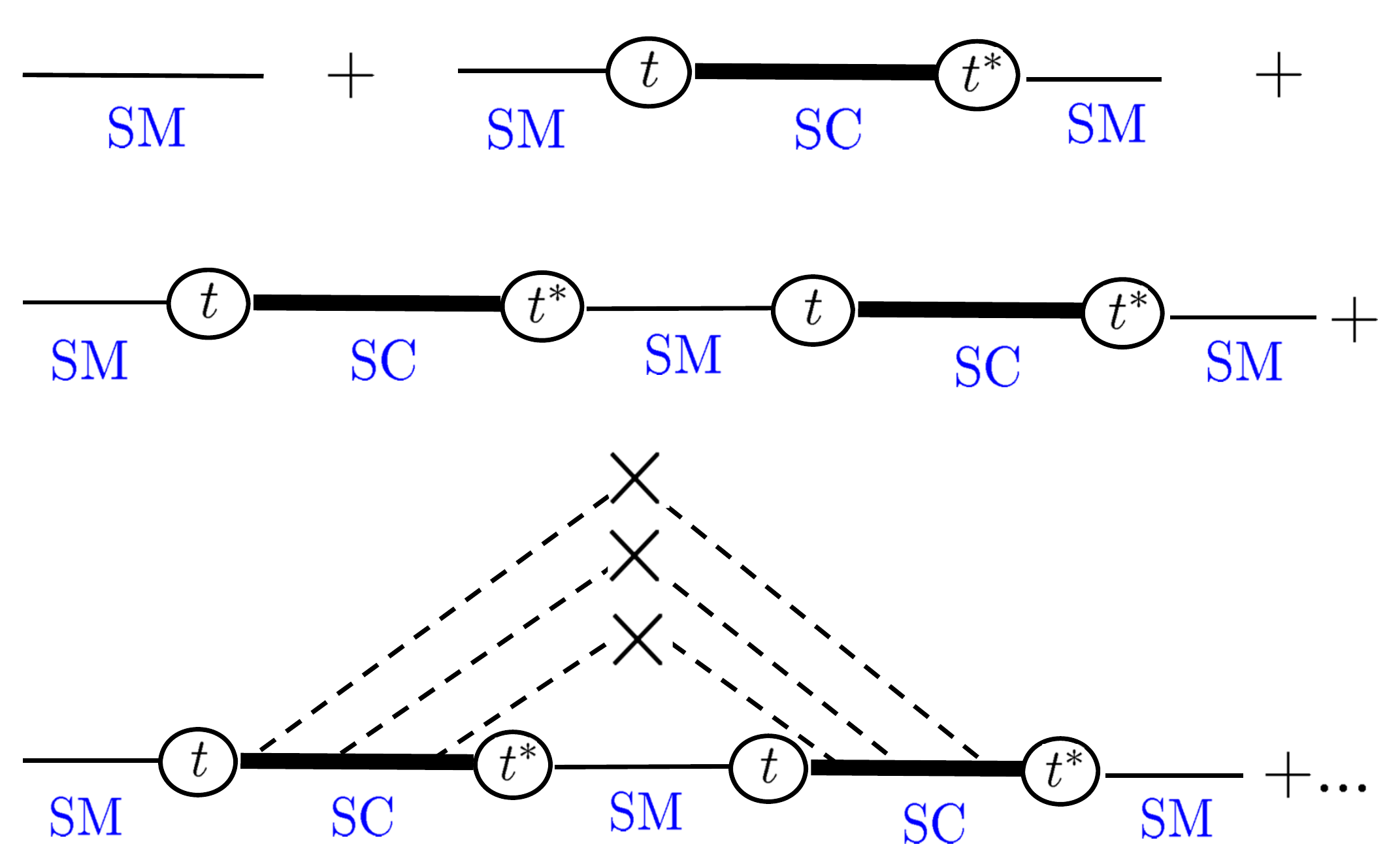}
\vspace{-7mm}
\end{center}
\caption{(Color online) Diagrammatic perturbation theory in the tunneling between semiconductor and superconductor. Disorder-averaging is performed at each order in tunneling $t$. The thick solid line represents disorder-averaged Green's function in the superconductor $\bar{G}(\bm p,\omega)$. The bottom diagram corresponds to irreducible contributions as far as disorder averaging is concerned and is calculated in the manuscript using exact eigenstates formalism.} \label{fig:disorder}
\end{figure}

The scattering rate of an electron with momentum $p$ can be calculated using Fermi's Golden rule:
\begin{align}
\Gamma=2\pi\sum_{p'}|A_{p,p'}|^2\delta(\xi_p-\xi_{p'})
\end{align}
with $A_{p,p'}$ being the amplitude for scattering to the state $p^\prime$, which can be calculated perturbatively in tunneling $t$
\begin{align}\label{eq:amplitude1}
A_{p,p'}\!=\!\bra{p,\sigma}H_t-H_t \frac{1}{H_{\rm S}} H_t + ... \ket{p',\sigma}.
\end{align}
It is easy to show that there is no contribution to the amplitude in the lowest order of perturbation theory (see also Ref.~\onlinecite{Stanescu'11}) and one has to consider higher order processes. The lowest non-zero contribution appears in the second order in $H_t$. After simple algebra, one finds that the amplitude in the second order in $|t|$ reads
\begin{align}\label{eq:amplitude2}
&A_{p,p'}\!=\!\frac{1}{S}\sum_{n,\sigma',\sigma''} \int\!\! d\bm x_1 d \bm r'_1 d\bm x_2 d \bm r'_2 T(\mathbf{x_1},\mathbf{r_1'}) T(\mathbf{x_2},\mathbf{r_2'})\\
\!&\!\times \!
\!\bra{\bm p,\sigma}\!\psi^\dag_{\sigma'}(\bm r_1)\!\ket{0}\!\!\bra{0}\!\psi_{\sigma''}(\bm r_2)\!\ket{\bm p',\sigma}
 \frac{\varepsilon_n}{\varepsilon_n^2\!+\!\Delta_0^2}\phi_n(\bm x_1)\phi_n(\bm x_2).\nonumber
\end{align}
The factor $\varepsilon_n/(\varepsilon_n^2+\Delta_0^2)$ appears due to the cancelation between particle and hole contributions to the amplitude $A_{p,p'}$. The amplitude~\eqref{eq:amplitude2} depends on eigenenergies $\varepsilon_n$ and eigenfunctions $\phi_n(\bm x)$ obtained for a particular realization of the disorder potential. Therefore, one needs to average the rate $\Gamma$ over different disorder realizations. The exact eigenstates formalism is very convenient here as we will show below. Alternatively and equivalently, one could do a diagrammatic calculation of the imaginary part of the self-energy by doing a perturbative expansion in tunneling and summing ladder diagrams due to the disorder as shown in Fig.~\ref{fig:disorder}. For s-wave superconductors these two approaches are equivalent~\cite{Ambegaokar'92}.

We proceed by first introducing the identity
$\sum_n \frac{\varepsilon_n}{\varepsilon_n^2+\Delta_0^2}\phi_n(\bm x_1)\phi_n(\bm x_2)=\int d \xi \frac{ \xi}{ \xi^2+\Delta_0^2} K(\xi,\bm x_1, \bm x_2)$, where $K(\xi,\bm x_1, \bm x_2) =\sum_n \delta( \xi-\varepsilon_n) \phi_n(\bm x_1)\phi_n(\bm x_2)$ and reducing disorder averaging of the rate $\Gamma$ to finding correlation functions of $K(\xi,\bm x_1, \bm x_2)$. After straightforward manipulations, the disorder-averaged scattering rate becomes
\begin{align}\label{eq:rate}
&\langle \Gamma \rangle= 2\pi\sum_{p'}\delta(\xi_p-\xi_{p'})\frac{1}{S^2}\int\!\! \prod_{i=1..4} d\bm x_i d \bm r'_i \\
& \times T(\mathbf{x_1},\mathbf{r_1'}) T(\mathbf{x_2},\mathbf{r_2'}) T(\mathbf{x_3},\mathbf{r_3'}) T(\mathbf{x_4},\mathbf{r_4'})e^{i\bm p (\bm r'_1\!-\!\bm r'_3)\!-\!i\bm p' (\bm r'_2\!-\!\bm r'_4)}\nonumber\\
&\times  \int d \xi'  \int d \xi \frac{ \xi}{ \xi^2+\Delta_0^2} \frac{ \xi'}{ \xi'^2+\Delta_0^2} \langle K_{\xi}(\bm x_1, \bm x_2) K_{\xi'}(\bm x_3,\bm x_4) \rangle. \nonumber
\end{align}
Here the brackets $\langle ... \rangle$ denote averaging
over different realizations of the random potential $\mathcal V(x)$ in the superconductor. The correlation function
$\langle K_{\xi_1}(\mathbf{x_1},\mathbf{x_2})
K_{\xi_2}(\mathbf{x_3},\mathbf{x_4})\rangle$ consists of
reducible and irreducible parts,
\begin{align}\label{KK2}
\langle
&K_{\xi_1}(\mathbf{x_1},\mathbf{x_2})K_{\xi_2}(\mathbf{x_3},\mathbf{x_4})
\rangle\!=\\
&\!\!\!\langle K_{\xi_1}(\mathbf{x_1},\mathbf{x_2})\rangle
\!\langle
K_{\xi_2}(\mathbf{x_3},\mathbf{x_4})\rangle
\!+\!\langle K_{\xi_1}(\mathbf{x_1},\mathbf{x_2})
K_{\xi_2}(\mathbf{x_3},\mathbf{x_4})\!\rangle_{\rm{ir}}.\nonumber
\end{align}
The reducible part can be easily calculated by relating $\langle
K_{\xi}(\mathbf{x_1},\mathbf{x_2})\rangle$ to the
ensemble-averaged normal-state Green's function: $\!\langle
K_{\xi}(\mathbf{x_1},\mathbf{x_2})\rangle\equiv\!-\frac{1}{\pi}{\rm{Im}}\langle
G^R_{\xi}(\mathbf{x_1},\mathbf{x_2})\rangle\!=\!\nu_F f_{12}$.
(Upon averaging over disorder, one can neglect the energy
dependence of the density of states here, \emph{i.e.}
$\langle\nu_F(\xi)\rangle=\nu_F$. The function $f_{12}$ is given
by $f_{12}=\langle
e^{i\mathbf{k}(\mathbf{\mathbf{x_1}}\!-\!\mathbf{x_2})}\rangle_{\rm{FS}}$
with $\langle...\rangle_{\rm{FS}}$ being the average over electron
momentum on the Fermi surface. For 3D systems the function $f_{12}$
is equal to
$f_{12}=~\frac{\sin(k_F|\mathbf{x_1}\!-\!\mathbf{x_2}|)}{k_F|\mathbf{x_1}\!-\!\mathbf{x_2}|}$.)
The irreducible part $\langle K_{\xi_1}(\mathbf{x_1},\mathbf{x_2})
K_{\xi_2}(\mathbf{x_3},\mathbf{x_4}) \rangle_{\rm{ir}}$ can be
expressed in terms of the classical diffusion propagators
 - diffusons and Cooperons, see, for example, Ref.~[\onlinecite{Aleiner2002}]. Assuming that time-reversal symmetry is preserved in the superconductor, diffusons and Cooperons coincide,
$\mathcal{P}_{\omega}(\mathbf{x}_1,\mathbf{x}_2)=\mathcal{P}^{D}_{\omega}(\mathbf{x}_1,\mathbf{x}_2)=\mathcal{P}^{C}_{\omega}(\mathbf{x}_1,\mathbf{x}_2)$,
and the irreducible part of the correlation function~(\ref{KK2})
reads
\begin{eqnarray}\label{KK3} \!\!\!\!\!\!&\,&\!\!\!\!\!\!\!\!\!\!\langle
K_{\xi_1}(\mathbf{x_1},\mathbf{x_2})
K_{\xi_2}(\mathbf{x_3},\mathbf{x_4})
\rangle_{\rm{ir}}=\\
\!\!&\!=\!&\frac{\nu_F}{\pi}\mbox{Re}\left[f_{14}f_{23}\mathcal{P}_{|\xi_2\!-\!\xi_1\!|}(\mathbf{x}_1,\mathbf{x}_3)\!+\!f_{13}f_{24}\mathcal{P}_{|\xi_2\!-\!\xi_1\!|}(\mathbf{x}_1,\mathbf{x}_4)\right].\nonumber
\end{eqnarray}
Using these results, we can now perform integrals over $\xi$ and $\xi'$ in Eq.~\eqref{eq:rate}.
Given that the reducible part of the correlation function is independent of energy, the energy integrals vanish. Thus, irreducible terms do not contribute to the momentum relaxation rate, see also diagrammatic calculation in Ref.~\onlinecite{Stanescu'11}. The contribution of the irreducible part is proportional to
\begin{align}\label{eq:diffusion}
&F(\mathbf x,\mathbf x')=\int d \xi'  \int d \xi \frac{ \xi}{ \xi^2+\Delta_0^2} \frac{ \xi'}{ \xi'^2+\Delta_0^2} \mathcal{P}_{|\xi_2\!-\!\xi_1\!|}(\mathbf{x},\mathbf{x'})\nonumber\\
&=\int d \xi'  \int d \xi \frac{ \xi}{ \xi^2+\Delta_0^2} \frac{ \xi'}{ \xi'^2+\Delta_0^2}\int dt e^{i(\xi-\xi')t}\mathcal{P}(t,\mathbf{x},\mathbf{x'})\nonumber\\
\!&=\!\pi^2\!\int\! dt\! e^{-2\Delta_0|t|}\mathcal{P}(t,\mathbf{x},\mathbf{x'})\!=\!\pi^2\frac{\exp[-\sqrt 6|\mathbf x-\mathbf x'|/\bar \xi]}{4\pi D |\mathbf x-\mathbf x'|},
\end{align}
where $\mathcal{P}(t,\mathbf{x},\mathbf{x'})$ is a solution of a 3D diffusion equation:
$\mathcal{P}(t,\mathbf{x},\mathbf{x'})=\exp\left(-\frac{|\bm x-\bm x'|}{4Dt}\right)/(4\pi Dt)^{3/2}$.
Here $D$ is the diffusion constant $D=l v_F/3$ with $l$ and $v_F$ being the mean-free path and Fermi velocity, respectively; $\bar\xi=\sqrt{\xi_0 l}$ is the effective coherence length in disordered superconductors and $\xi_0=v_F/\Delta_0$.  Thus, the expression for the rate $\langle \Gamma \rangle$~\eqref{eq:rate} now becomes
\begin{align}
&\langle \Gamma \rangle \!=\! \frac{2\pi}{S^2}\sum_{p'}\delta(\xi_p\!-\!\xi_{p'})\int\!\! \prod_{i=1..4} d\bm x_i d \bm r'_i e^{i\bm p (\bm r'_1-\bm r'_3)-i\bm p' (\bm r'_2-\bm r'_4)}\nonumber\\
&\times T(\mathbf{x_1},\mathbf{r_1'}) T(\mathbf{x_2},\mathbf{r_2'}) T(\mathbf{x_3},\mathbf{r_3'}) T(\mathbf{x_4},\mathbf{r_4'})\nonumber\\
&\times \frac{\nu_F}{\pi}\left[f_{14}f_{23}F(\mathbf{x}_1,\mathbf{x}_3)\!+\!f_{13}f_{24}F(\mathbf{x}_1,\mathbf{x}_4)\right].
\end{align}
Taking into account Eq.~\eqref{eq:tunn}, one can now compute the spatial integrals. The integrand is a quickly decaying function and converges at short length scales $~p_F^{-1}$ where $p_F$ is the Fermi momentum in the superconductor. Thus, the dominant contribution comes from $x_1\approx x_4$, $x_2\approx x_3$ and $x_1\approx x_3$, $x_2\approx x_4$, respectively, and one finds that $\langle \Gamma \rangle=\langle \Gamma_1 \rangle+ \langle \Gamma_2 \rangle$:
\begin{align}
\!\langle  \Gamma_1 \rangle \!&\!=\!\sum_{p'}\!\delta(\!\xi_p\!-\!\xi_{p'}\!)\!\!\frac{(2\pi)^3 |t|^4\!\nu_F}{\pi S^2}\!\!\int\!\! d\!\bm r_1 d\bm r_2 \!\!e^{i(\bm p\!+\!\bm p')\!(\bm r_1\!-\!\bm r_2)} \!F(\mathbf{r}_1,\!\mathbf{r}_2)\nonumber\\
\!&\!=\! 2\pi\sum_{p'}\!\delta(\xi_p-\xi_{p'})\!\frac{2\pi^3 |t|^4 \nu_F}{S D}\frac{\bar \xi}{\sqrt{\bar \xi^2 |\bm p+\bm p'|^2+6}}\nonumber\\
& \approx
\left \{\begin{array}{c}
    \frac{12 \pi^2 |t|^4 \nu_F m^*}{ \sqrt 2 v_F l p^{\rm SM}_F}\log[p^{\rm SM}_F\bar \xi] \mbox{ for $p^{\rm SM}_F \bar\xi \gg 1$} \\ \, \\
    \frac{\sqrt 6\pi^3 |t|^4 \nu_F m^*\bar \xi}{v_F l}  \mbox{ for $p^{\rm SM}_F \bar\xi \ll 1$}
  \end{array} \right.
 \\
\langle \Gamma_2 \rangle &= 2\pi\sum_{p'}\delta(\xi_p-\xi_{p'})\frac{(2\pi)^2 |t|^4 \nu_F}{\pi S^2}\int\!\! d^2 \bm r_1 d^2 \bm r_2 F(\mathbf{r}_1,\mathbf{r}_2)\nonumber\\
&=2\pi\sum_{p'}\delta(\xi_p-\xi_{p'})\frac{2\pi^3 |t|^4 \nu_F \bar \xi}{\sqrt 6 D S}\nonumber\\
&=\frac{\sqrt 6\pi^3 |t|^4 \nu_F m^*\bar \xi}{v_F l}
\end{align}
Here $p^{\rm SM}_F$ is the Fermi momentum in the semiconductor, and we have assumed that $p_F\bar \xi \gg 1$.
It is convenient to re-write the above expressions for the scattering rate $\langle \Gamma \rangle$ in terms of the level broadening $\gamma=2\pi \nu_F |t|^2$ in the semiconductor due to the presence of a bulk metal. Then,  one can estimate the upper bound on $\langle \Gamma \rangle$ to be
\begin{align}\label{eq:ratefinal}
\langle \Gamma \rangle \approx \sqrt 6 \pi^3 \frac{\gamma^2}{\Delta_0}\frac{m^*}{m}\frac{1}{p_F \bar \xi}.
\end{align}
As follows from Eq.~\eqref{eq:ratefinal}, the momentum relaxation rate due to the presence of a disordered superconductor is proportional to $\gamma^2$ whereas the proximity-induced gap is of the order of $\gamma$. Furthermore, $\langle \Gamma \rangle$ is additionally suppressed due to quantum interference effects by a nontrivial (and non-obvious) factor $1/p_F\bar\xi \ll 1$.

Our results for the momentum relaxation rate can be qualitatively explained as follows. From Eq.~\eqref{eq:diffusion}, one can see that the rate $\langle \Gamma \rangle$ is proportional to the probability of a quasiparticle to return to the junction within the time $\Delta_0^{-1}$ which is the time an unpaired quasiparticle can spend in the superconductor in a virtual state. This introduces a length scale in the problem $\bar\xi \propto \sqrt {D/\Delta_0}$ above which return probability is exponentially suppressed. Therefore, one can think that the effective size of the system relevant for this process is of the order $\bar\xi$.
The momentum relaxation rate generated by a superconducting layer of size $\bar \xi$
is proportional to the attempt frequency $\gamma_{\rm in}\sim |t|^2 \nu_F$ times the probability to return to the semiconductor within time $\Delta_0^{-1}$.
The latter is given by $\gamma_{\rm out}/\Delta_0 \ll 1$ where $\gamma_{\rm out}\sim |t|^2 \nu^{\rm 2D}_F/\bar{\xi}$. (Here $|t|^2/\bar\xi^3$ has dimension $E^2$ and $\nu^{\rm 2D}_F \bar\xi^2$ scales as $1/E$). Combining all the terms, we finally recover the expression for the rate $\langle \Gamma \rangle$ given by Eq.~\eqref{eq:ratefinal}. In particular, these arguments explain why the rate is suppressed by a factor $1/p_F\bar\xi$.

We now take into account the effect of Rashba spin-orbit coupling and Zeeman splitting in the semiconductor and discuss momentum relaxation rate in the helical phase. The situation we are interested in is when the chemical potential is in the gap as shown in Fig.~\ref{fig:system}a. Thus, the momentum relaxation rate is determined by the scattering amplitude between different momenta on the helical Fermi surface which is given by $A_{p,p'}\!=\!-\bra{p,-} H_t \frac{1}{H_{\rm S}} H_t \ket{p',-}$. Here $\ket{p, \pm}$ denotes the state of an electron on the Fermi surface with a particular chirality $\pm$. The results of the calculation can be straightforwardly obtained by repeating the steps above but are not particular illuminating.
For experimentally relevant parameters,
the rate $\langle \Gamma \rangle$ has to be multiplied by a factor $O(1)$ which originates from the change in the matrix elements, as well as the density of states. Therefore, for practical purposes one can use Eq.~\ref{eq:ratefinal} in 2D system. It is worth pointing out, however, that in the limit $m^*\alpha^2\gg V_z$ there is an additional suppression of the impurity scattering rate because the Berry phase of the Fermi surface is equal to $\pi$ up to corrections $V_z/m^*\alpha^2 \ll 1$. This suppression of the elastic backscattering is particularly important for one-dimensional helical nanowires. Overall, the rate \eqref{eq:ratefinal} should be considered as an upper bound on the effective impurity scattering rate due to superconducting disorder. We can now estimate $\langle \Gamma \rangle/\Delta_{\rm ind}$. Taking InAs as a semiconductor and Al (or Nb) as a superconductor, we find that $1/p_F\bar\xi \sim 10^{-3}$ and $m^*/m\approx 0.04$. The effective proximity-induced gap varies from $\Delta_{\rm ind}\sim \gamma$ for $m^*\alpha^2\gg V_z$ and $\Delta\sim \gamma \sqrt{m^*\alpha^2/V_z}$ for $m^*\alpha^2\ll V_z$. Assuming that $\gamma \sim \Delta_0$ and taking the pessimistic numbers for spin-orbit coupling $m^*\alpha^2/V_z\sim 0.1$, one finds that the ratio of the scattering rate $\langle \Gamma \rangle$~\eqref{eq:ratefinal} to the proximity-induced gap $\Delta_{\rm ind}\sim\gamma$ is small $\langle \Gamma \rangle/\Delta_{\rm ind}\sim 10^{-2}$. This ratio involving the upper bound of the momentum relaxation rate is thus very small and we, therefore, conclude that the topological superconducting phase emerging at the interface is robust against disorder in the superconductor.


We would like to thank Leonid Glazman, Patrick Lee,  Chetan Nayak, Felix von Oppen and Andrew Potter for stimulating discussions. This work is supported by the DARPA QuEST, JQI-NSF-PFC, Microsoft Q (SDS) and WVU startup funds (TS). RL thanks the Aspen Center for Physics for hospitality and support under the NSF grant \#1066293.


%

\end{document}